%% file: root.tex
\documentclass[letterpaper, 10 pt, conference]{ieeeconf}  

\IEEEoverridecommandlockouts                              

\usepackage{amsmath,amsfonts}
\usepackage[ruled,vlined]{algorithm2e}

\usepackage{array}
\usepackage[caption=false,font=normalsize,labelfont=sf,textfont=sf]{subfig}
\usepackage{textcomp}
\usepackage{stfloats}
\usepackage{float}
\usepackage{url}
\usepackage{verbatim}
\usepackage{graphicx}
\usepackage{cite}
\usepackage{mathtools} 
\usepackage{subcaption}
\usepackage{amssymb}  

\usepackage{amsthm}
\usepackage{algpseudocode}
\usepackage{xcolor}
\usepackage{accents}
\usepackage{mathrsfs}
\usepackage[normalem]{ulem}

\input{definitions}

\usepackage{soul}
\usepackage{booktabs}
\usepackage{amsmath} 

\usepackage{multirow}

\usepackage[T1]{fontenc}
\usepackage[utf8]{inputenc}
\usepackage[skip=0.333\baselineskip]{caption}
\usepackage{siunitx} 
\newcolumntype{T}{S[table-format=3.3, input-symbols={()},
                    table-space-text-post={$^{***}$},
                    table-align-text-post=false]}
\usepackage[colorlinks=true, allcolors=black]{hyperref}
\usepackage{tabularx,booktabs}
\newcolumntype{C}{>{\centering\arraybackslash}X} 
\newtheorem{theorem}{Theorem}

\newtheorem{definition}{Definition}

\definecolor{green}{RGB}{11,155,13}

\SetCommentSty{mycommfont}

\SetKwRepeat{Do}{do}{while}

\usepackage{yaacro}

\begin{acgroupdef}[list=acronimos]
    \acdef{JSG}{Joint State Graph}
\end{acgroupdef}

\newcommand{\nodeset}{\mathbb{V}}
\newcommand{\edgeset}{\mathbb{E}}

\title{\LARGE \bf
Team Coordination on Graphs: Problem, Analysis, and Algorithms
}

\author{Yanlin Zhou, Manshi Limbu, Gregory J. Stein, Xuan Wang, Daigo Shishika, and Xuesu Xiao
\thanks{George Mason University {\tt\scriptsize \{yzhou30, klimbu2, gjstein, xwang64, dshishik, xiao\}@gmu.edu}. 
This work has taken place in the RobotiXX Laboratory at George Mason University. RobotiXX research is supported by Army Research Office (ARO, W911NF2220242, W911NF2320004, W911NF2420027), US Air Forces Central (AFCENT), Google DeepMind (GDM), Clearpath Robotics, and Raytheon Technologies (RTX).
}
}

\begin{document}
\maketitle
\thispagestyle{empty}
\pagestyle{empty}

\input{Contents/abstract}
\input{Contents/intro}
\input{Contents/related}

\input{Contents/method}

\input{Contents/analysis}

\input{Contents/solution}

\input{Contents/results}
\input{Contents/conclusions}

\bibliographystyle{ieeetr}
\bibliography{bib}
\end{document}

%% file: definitions.tex
\def\beq{\begin{equation*}}
\def\eeq{\end{equation*}}
\def\bql{\begin{equation}}
\def\eql{\end{equation}}
\def\bqn{\begin{eqnarray*}}
\def\eqn{\end{eqnarray*}}
\def\bnl{\begin{eqnarray}}
\def\enl{\end{eqnarray}}



%% file: Contents/abstract.tex

\begin{abstract}
    Team Coordination on Graphs with Risky Edges (\textsc{tcgre}) is a recently emerged problem, in which a robot team collectively reduces graph traversal cost through support from one robot to another when the latter traverses a risky edge. 
    Resembling the traditional Multi-Agent Path Finding (\textsc{mapf}) problem, both classical and learning-based methods have been proposed to solve \textsc{tcgre}, however,  they lacked either computational efficiency or optimality assurance. 
    In this paper, we reformulate \textsc{tcgre} as a constrained optimization problem and perform a rigorous mathematical analysis.
    Our theoretical analysis shows the NP-hardness of \textsc{tcgre} by reduction from the Maximum 3D Matching problem and that efficient decomposition is a key to tackle this combinatorial optimization problem. 
    Furthermore, we design three classes of algorithms to solve \textsc{tcgre}, \textit{i.e.}, Joint State Graph (\textsc{jsg}) based, coordination based, and receding-horizon sub-team based solutions. Each of these proposed algorithms enjoys different provable optimality and efficiency characteristics that are demonstrated in our extensive experiments. 
\end{abstract}


%% file: Contents/intro.tex
\section{Introduction}
Multi-Agent Path Finding (\textsc{mapf}) is a trending problem in robotics~\cite{adler2002cooperative,hu2021distributed,yu2022surprising,surynek2010optimization,foerster2017stabilising,gupta2017cooperative}, as it lies in the core of many robotic applications, e.g., drone swarm control~\cite{hu2021distributed}, autonomous delivery~\cite{liu2019task}, and public transportation scheduling~\cite{adler2002cooperative}.  
Adding the possibility of team coordination~\cite{yan2013survey} between robots to \textsc{mapf} makes the problem more difficult. The need for coordination behaviors on large-scale multi-robot planning problems may exceed the computational capability of a centralized planner~\cite{luna2011efficient}, giving rise to decentralized planning that distributes the computation into each robot~\cite{capitan2013decentralized}. 
However, despite the efficiency, dexterity, and responsiveness, the distribution itself may induce certain performance degradation and lead to suboptimal team coordination behaviors. Therefore, some centralized pre-planning is still crucial for multi-robot planning, especially for large-scale problems.

Team Coordination on Graphs with Risky Edges (\textsc{tcgre})~\cite{limbu2023team}  is such a centralized planning problem in an environment represented as a graph---multiple robots travel from their start to goal nodes with possible support from some nodes to reduce the cost of traversing certain risky edges, requiring team coordination behaviors to reduce the total cost of team graph traversal. By converting the environment graph to a Joint State Graph (\textsc{jsg}), optimal coordination can be solved using Dijkstra's search algorithm on the \textsc{jsg}~\cite{limbu2023team}. However, the conversion to \textsc{jsg} does not scale well with large environment graphs and number of robots. To address the curse of dimensionality, a Critical Joint State Graph (\textsc{cjsg}) approach has been proposed for large graphs with a small amount of support with up to two robots, still assuring solution optimality. Reinforcement Learning (RL) has been utilized~\cite{limbuteam} to reduce the time complexity and scale the solution to a large group of robots and the size of the graph, but at the cost of sacrificing optimality. 

To acquire theoretical insights into this problem, we reformulate \textsc{tcgre} in a constrained optimization framework and present a rigorous mathematical analysis of this reformulated problem. We prove the NP-hardness of \textsc{tcgre} by reduction from the Maximum 3D Matching problem. We further show that such a difficult combinatorial optimization problem can be effectively addressed by efficient decomposition. In addition to providing a theoretical explanation for previous  algorithms, we further introduce three distinct classes of methods to solve the \textsc{tcgre} problem: (1) Based on the idea of \textsc{jsg}, we introduce new search algorithms that do not need to fully construct the \textsc{jsg} in advance and can be guided by a new admissible heuristic, while guaranteeing optimal solutions; 
(2) Inspired by the Conflict-Based Search~\cite{sharon2015conflict}, the second class of algorithms is based on coordination and we design a Coordination-Exhaustive Search (\textsc{ces}) algorithm. \textsc{ces} starts with individual optimal paths and finds the lowest cost among every possible coordination for every robot to achieve the optimal solution within polynomial time with respect to the number of robots, under the assumption that the coordination between every pair of support node and risky edge is only necessary for a limited number of times;
(3) Motivated by \textsc{ces}'s assumption, we also propose a class of receding-horizon sub-team solutions that further decomposes the order of coordination by only looking at sub-team coordinations in a local region. We design a Receding-Horizon Optimistic Cooperative A* (\textsc{rhoc-a*}) search, in order to reduce the time complexity without much performance loss. 
Extensive experimental results are presented and discussed to inform the best ways to solve different \textsc{tcgre} problems.  

%% file: Contents/related.tex
\section{Related Work}
\label{sec::related}

We first review related work on the classical \textsc{mapf} problem and common classes of algorithms. We then review previous approaches to solve the \textsc{tcgre} problem. 

\subsection{\textsc{mapf} and Classes of Algorithms}

\textsc{mapf} is a specific type of multi-agent planning problem with a key constraint that no agents can collide with one another~\cite{stern2019multi}. 
A feasible solution to the problem is a joint plan that allows all agents to reach their goals from their starts. Two common objectives are makespan and total cost. Classical \textsc{mapf} problems may include additional assumptions, such as no vertex conflict, no edge conflict, no cycle conflict, and no swapping conflict~\cite{standley2010finding,felner2017search}. 

Algorithms to solve \textsc{mapf} include A*-based search with exponential space and time compelexity~\cite{ryan2008exploiting,standley2010finding}; conflict-based search~\cite{sharon2015conflict} by decomposing into many constrained single-agent problems; reduction-based approaches to SAT~\cite{surynek2012towards,surynek2016makespan}, ILP~\cite{yu2013planning}, ASP~\cite{erdem2013general}, or CSP~\cite{surynek2016efficient,bartak2017modeling}; rule-based algorithms based on Kornhauser’s algorithm~\cite{Kornhauser1984}, Push-and-Rotate~\cite{de2014push}, or BIBOX~\cite{surynek2009novel}; and suboptimal solutions~\cite{holte1996hierarchical, korf1990real} that sacrifice optimality for efficiency. 

\textsc{mapf} is NP-hard~\cite{goldreich2011finding}, and no optimal solutions can be found in polynomial time. The time complexity of all above optimal algorithms~\cite{ryan2008exploiting,standley2010finding,surynek2012towards,surynek2016makespan,sharon2015conflict,yu2013planning,erdem2013general,surynek2016efficient,bartak2017modeling} is  exponential to the number of agents. 
Similarly, we prove in this paper that our \textsc{tcgre} problem that utilizes, instead of avoiding, interactions between agents in the form of support is also NP-hard.

\subsection{Team Coordination on Graphs with Risky Edges (\textsc{tcgre})}
\textsc{tcgre}~\cite{limbu2023team} is a recently proposed problem, in which a team of robots traverses a graph from their starts to goals and supports each other while traversing certain risky (high-cost) edges to reduce overall cost. Instead of focusing on collision-free paths in the traditional \textsc{mapf}, the \textsc{tcgre} problem pursues team coordination. 
To solve \textsc{tcgre}, Limbu et al.~\cite{limbu2023team} have proposed \textsc{jsg} and \textsc{cjsg}, both of which construct a single-agent joint-state graph. After the construction, the original team coordination problem can be solved using Dijkstra's algorithm to solve a shortest path problem with optimality guarantee. The \textsc{cjsg} construction deals with the team coordination problem more efficiently, although it can only solve problems with two agents. 
To scale up \textsc{tcgre}, RL~\cite{limbu2023team} has been utilized to handle many nodes and robots, but at the cost of optimality.  

In this work, we reformulate the \textsc{tcgre} in a constrained optimization framework and conduct a mathematical analysis of this problem. We prove its NP-hardness and point out the necessity of efficient decomposition to effectively solve this problem. We further present three classes of algorithms to solve \textsc{tcgre} from different perspectives.

%% file: Contents/method.tex
\section{Problem Formulation}
\label{sec::method}


Assuming a team of $N$ homogeneous robots traverses an undirected graph \(\mathbb{G}=(\nodeset,\edgeset)\), where \(\nodeset\) is the set of nodes the robots can traverse to and \(\edgeset\) is the set of edges connecting the nodes, i.e., \(\edgeset\subset \nodeset\times \nodeset\). 
The team of robots traverses in the graph from their start nodes $\nodeset_0 \subset \nodeset$ to goal nodes $\nodeset_g \subset \nodeset$ via edges in \(\edgeset\). Each edge $e_{ij}=(V_i,V_j)\in \edgeset$ is associated with a cost $c_{ij}$, depending on its length, condition, traffic, obstacles, etc. Specially, some edges with high costs are difficult to traverse through, denoted as risky edges $\edgeset' \subset \edgeset$, but with the support from a teammate from a supporting node, their costs can be significantly reduced to $\tilde{c}_{ij}$. 
In this problem, we only consider such coordination behaviors between two robots. In one coordination behavior, one receiving robot receives support while traversing a risky edge, and another supporting robot offers support from some (nearby) location, called support node. Note that each risky edge $e_{ij} \in \edgeset'$ corresponds to certain support node(s) $\mathbb{S}_{e_{ij}} \subset \nodeset$ ($\mathbb{S}_{e_{ij}}=\emptyset$ if $e_{ij}\notin \edgeset'$). Additionally, the coordination also induces some cost for the supporter, denoted by $c'$.
A central planner needs to schedule the paths of all agents and coordination on their ways.

\subsection{Action \& Cost Model}
\label{CostModel}
Without coordination,
at each time step $t$, a robot $n$ can choose to stay where it is, or move to its neighbor ($V_i$ is the neighbor of $V_j$ if $e_{ij}\in \edgeset$). Its movement can be denoted by $M_n^t=(l_n^t,l_{n}^{t+1}) \in \edgeset$, where $l_n^t, l_{n}^{t+1} \in \nodeset$ indicate its current and next location and $l_{n}^{t+1}$ is a neighbor of $l_n^t$. Specially, the robot stays at its current location if $l_{n}^{t+1}=l_n^t$ with zero cost, i.e., $c_{ii}=0,\forall i$. 
The movement set can thus be denoted by $\mathcal{M} = \{M_{n}^{t}|\forall n, \forall t\}$. 
Moreover, the movement decision $M_{n}^{t}$ can be rewritten as an 0/1 variable $M_{ij}^{nt}$, where $M_{ij}^{nt}=1$ represents edge $e_{ij}$ is selected by robot $n$ at time $t$, and 0 otherwise. A robot can only move once at each time step to a neighbor node or not at all, i.e., $\sum_{\forall e_{ij}\in \mathcal{N}_{l_n^t}}M_{ij}^{nt}=1 ~\&~ \sum_{\forall e_{ij}\notin \mathcal{N}_{l_n^t}}M_{ij}^{nt}=0$, where $\mathcal{N}_{l_n^t}=\{(l_n^t,l_n^{t+1})|\forall (l_n^t,l_n^{t+1})\in \edgeset\}$. The movement set can thus be denoted by $\mathcal{M} = \{M_{ij}^{nt}| \forall i,j,\forall n, \forall t\}$. 

When a coordination behavior is available---when robot $n$ is going to traverse a risky edge, another robot $m$ happens to be at one of the support nodes of the risky edge or vice versa, i.e., $M_n^t\in \edgeset'$ and $l_m^t\in \mathbb{S}_{M_n^t}$, or $M_m^t\in \edgeset'$ and $l_n^t\in \mathbb{S}_{M_m^t}$---the robot pair needs to decide whether to provide/receive support. 
Denote the coordination decision of agent $n$ at time $t$ as $s_{nm}^t$. It is clear that agent $n$'s coordination decision is dependent on its movement decision, so the cost is twofold:
\noindent (1) When agent $n$ has no coordination opportunity (the above coordination behavior is not available for any other robot $m$), i.e., $\forall m$, $l_m^t\notin \mathbb{S}_{M_n^t}$ and $l_n^t\notin \mathbb{S}_{M_m^t}$, its cost $C_n^t$ is only decided by its movement, i.e., $C_n^t = c_{ij}$, where $M_{ij}^{nt} = 1$. 
 
\noindent (2) When coordination is possible for agent $n$, i.e., $\exists m$, $l_m^t\in \mathbb{S}_{M_n^t}$ or $l_n^t\in \mathbb{S}_{M_m^t}$, the cost $C_n^t$ can be represented as
\vspace{-5pt}
\begin{align}
\label{IndividualCost}
    C_n^t=\begin{cases}
        c_{ij}, ~~\textit{if} ~~ s_{nm}^t=~~0;\\
        \tilde{c}_{ij}~~~\textit{if} ~~ s_{nm}^t=~~1;\\
        c',~~~\textit{if} ~~ s_{nm}^t=-1.
    \end{cases}
\end{align}

\vspace{-5pt}
\noindent where $s_{nm}^t=1$ means agent $n$ decides to receive support from $m$,  $s_{nm}^t=-1$ indicates agent $n$ decides to offer support to $m$, and $s_{nm}^t=0$ stands for no coordination between the robot pair $n$ and $m$ at $t$.
Specially, no coordination happens for one single robot, i.e., $s_{nn}^t=0,\forall n,\forall t$, or when $n$ and $m$ cannot support each other, i.e., $s_{nm}^t=0$ if $l_m^t\notin \mathbb{S}_{M_n^t}$ and $l_n^t\notin \mathbb{S}_{M_m^t}$, $\forall m$. 
The coordination decision set can be written as $\mathcal{S}=\{s_{nm}^t|\forall n,m, \forall t\}$. In addition, a coordination decision is made for a pair, so $s_{nm}^t+s_{mn}^t=0$ for every pair of robots. Furthermore, the robots can wait now (no movement) for future coordination, but there is no point for all robots to stay still at the same time, i.e.,  $\sum_{\forall n}\sum_{\forall i\neq j} M_{ij}^{nt}\neq0$.

\subsection{Problem Definition}
\label{definition}
Given the node set $\nodeset$, the edge set $\edgeset$, support nodes for each edge $\mathbb{S}_{e_{ij}}$, cost of each edge without and with coordination $c_{ij},\tilde{c}_{ij}$, $N$ robots with their starts $\nodeset_0$ and goals $ \nodeset_g$, optimize the movement and coordination decisions $\mathcal{M}$ and $\mathcal{S}$, in order to minimize the total cost for each agent to traverse from its start to its goal within a time limit $T$.
Formally, the problem can be represented as
\begin{gather}
    \min_{{\mathcal{M}},{\mathcal{S}}} \sum_{t=0}^{T-1} \sum_{n=1}^N C_{n}^t. \label{objective}\\
    s.t.~~
     \sum_{\forall m \in \{1,2,...,N\}}|s_{nm}^t|\leq 1,\nonumber\\
    \forall n\in \{1,2,...,N\}, \forall t\in \{0,1,...,T-1\}. \label{constraint1:supportnum}\\
    s_{nm}^t, s_{mn}^t\in\{-1, 0,1\}, s_{nm}^t+ s_{mn}^t=0, \nonumber\\
    \forall n,m\in \{1,2,...,N\}, \forall t\in \{0,1,...,T-1\}. \label{constraint2:supportpair}\\
    l_n^0=\nodeset_0(n), \forall n\in \{1,2,...,N\}.\label{constraint3:starts}\\
    l_n^T=\nodeset_g(n), \forall n\in \{1,2,...,N\}.\label{constraint4:goals}\\
    \sum_{\forall e_{ij}\in{\mathcal{N}_{l_n^t}}}M_{ij}^{nt}=1,  \sum_{\forall e_{ij}\notin{\mathcal{N}_{l_n^t}}}M_{ij}^{nt}=0,                   \nonumber\\
    \forall n\in \{1,2,...,N\}, \forall t\in \{0,1,...,T-1\}.\label{constraint5:movenum}\\
    \sum_{\forall n}\sum_{\forall i\neq j} M_{ij}^{nt}\neq0,\nonumber\\
    \forall t\in\{0,1,...,T-1\}.\label{constraint6:correlation}
\end{gather}

\vspace{-5pt}
\noindent 
Eqn.~\eqref{objective} suggests the goal of the problem is to minimize the total cost of all agents across all time steps with two decision variables, movement set $\mathcal{M}$ and coordination set $\mathcal{S}$. Eqn.~\eqref{constraint1:supportnum} indicates that, at each time step, each robot can participate in at most one coordination behavior. Eqn.~\eqref{constraint2:supportpair} regulates that, at each time step, one coordination behavior only occurs between one robot pair. Eqn.~\eqref{constraint3:starts} and Eqn.~\eqref{constraint4:goals} set the start and the goal for each robot. Eqn.~\eqref{constraint5:movenum} guarantees that, at each time step, a robot can only move to a neighbor node or stay still. Eqn.~\eqref{constraint6:correlation} assures no unnecessary stagnation.

%% file: Contents/analysis.tex
\section{Mathematical Analysis}
\label{sec::sol}
\begin{figure*}[ht]
    \begin{minipage}[t]{0.3\linewidth}
    \centering
    \raisebox{0.13\height}{\includegraphics[width=0.99 \textwidth]{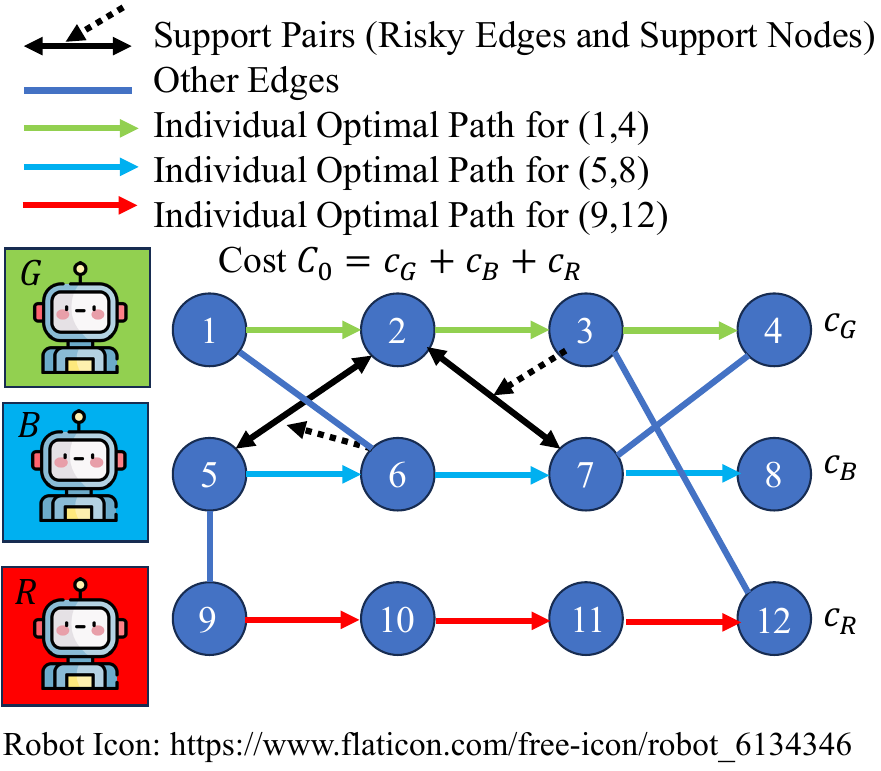}}
    \label{IndividualPaths}
    \end{minipage}
    \begin{minipage}[t]{0.39\linewidth}
    \centering
    \includegraphics[width=0.99\textwidth]{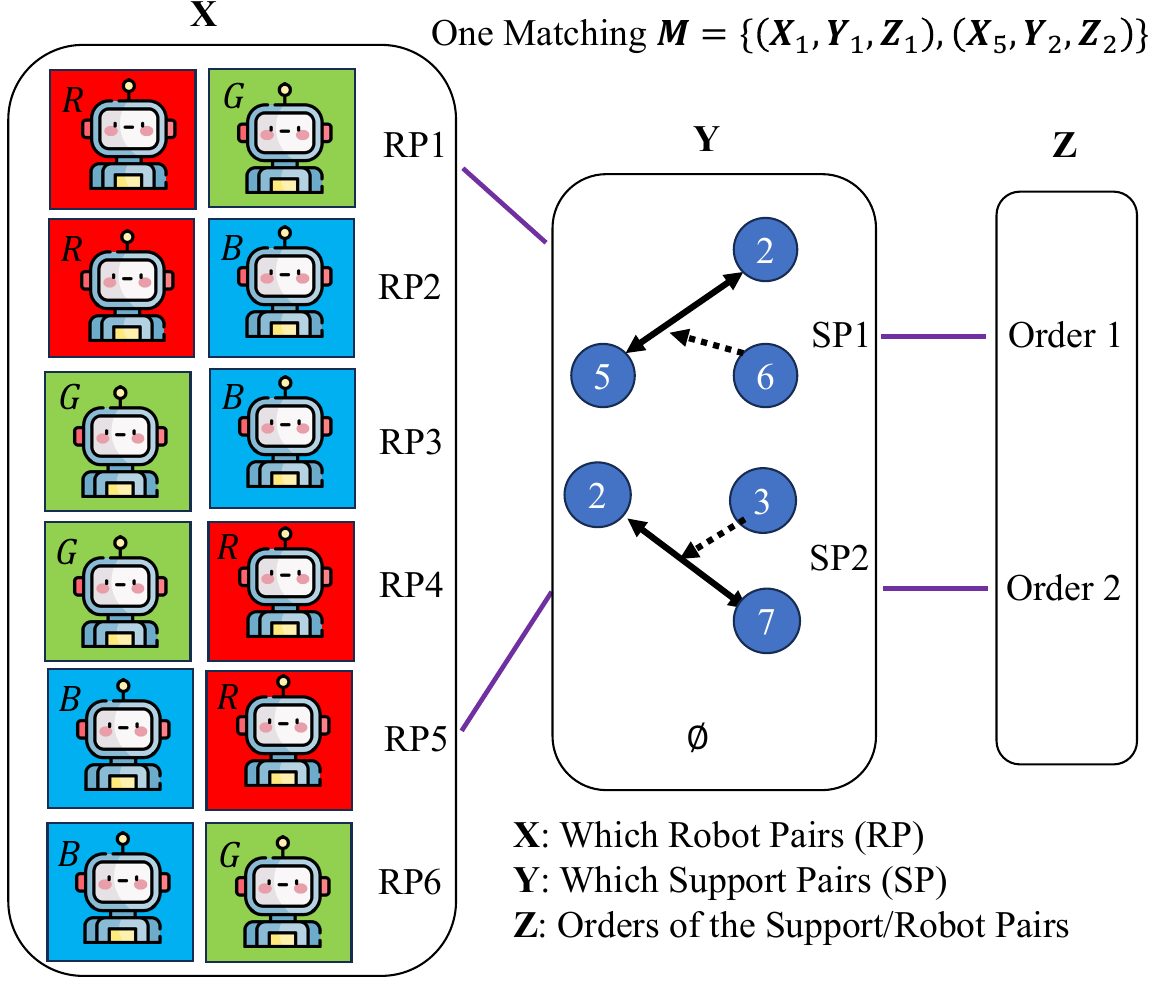}
    \label{3DMatchingGraph}
    \end{minipage}
    \begin{minipage}[t]{0.3\linewidth}
    \centering
    \raisebox{0.23\height}{\includegraphics[width=0.99 \textwidth]{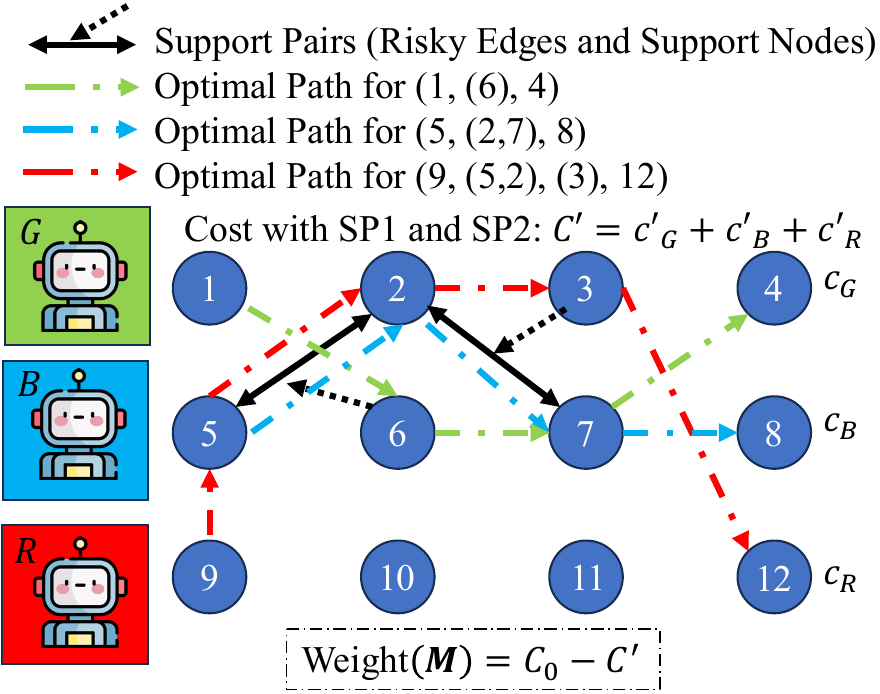}}
    \label{ReducedCost}
    \end{minipage}
\vspace{-10pt}
\caption{Reduction from Maximum 3D Matching (Middle) to \textsc{tcgre} (Left and Right) and Inspiration for \textsc{ces}. }
\label{Proof1}
\vspace{-10pt}
\end{figure*}
In this section, we prove our \textsc{tcgre} problem reduces from the Maximum 3D Matching problem, an NP-hard problem. Then, we start a rigorous analysis on the mathematical problem, which suggests decomposition is a promising solution to this combinatorial optimization problem. 

\subsection{NP-Hardness}

\begin{definition}
\label{3DMatching}
    Maximum 3D Matching: ${\bf{X}},{\bf{Y}},{\bf{Z}}$ are 3 finite sets. ${\bf{T}}$ is the subset of ${\bf{X}}*{\bf{Y}}*{\bf{Z}}$, with triples $(x,y,z)$, where $x\in {\bf{X}},y\in{\bf{Y}},z\in{\bf{Z}}$. ${\bf{M}}\subset  {\bf{T}}$ is a 3D matching if for any two distinct triples $(x_1, y_1, z_1)$ and $(x_2, y_2, z_2) \in {\bf{M}}$, we have $x_1 \neq x_2$, $y_1 \neq y_2$, and $z_1 \neq z_2$; each triple has a weight $w(x_i,y_j,z_k)$. Maximum 3D Matching problem is 
    to find a 3D matching with maximum total weight.
\end{definition}
\begin{theorem}
Maximum 3D Matching reduces to \textsc{tcgre}.
\end{theorem}
\vspace{-3pt}
\noindent\textit{Proof sketch:} Without coordination, we can generate individual optimal paths for all robots; the cost is $C_0$ (Fig.~\ref{Proof1} left). With coordination, we let some robot pairs take detours to the support pairs (i.e., one risky edge and one support node); the new cost is $C'$, where the first robot in the pair traverses the risky edge while the second visits and supports from the support node (Fig.~\ref{Proof1} right). Minimizing total cost is equivalent to maximizing the cost reduction $C_0-C'$.

    Consider {\bf{X}} contains all robot pairs; ${\bf{Y}}$ is the set of all support pairs plus an empty set element; {\bf{Z}} is a list of time orders of events (time steps are not necessarily needed, because robots can stay still and wait, Fig.~\ref{Proof1} middle). 
    Consider the weight $w(x_i,y_j,z_k)$ as the sum of cost reduction through coordination for robot pair $x_i$ to detour to support pair $y_j$ with time order $z_k$. The weight of a 3D matching $\textbf{M}$ is the total cost reduction of all robots, $C_0-C'$. Maximum 3D Matching is to find the paths for all robot pairs to achieve maximum total cost reduction from their original without-coordination costs, which is a specific case of the general \textsc{tcgre} problem that only needs to use each robot and support pair once with a smaller solution space than \textsc{tcgre}. 
    
    Since \textsc{tcgre} reduces from Maximum 3D Matching, a classical NP-hard problem~\cite{kann1991maximum}, \textsc{tcgre} is also NP-hard. We cannot find an optimal solution in polynomial time.\hfill \qedsymbol{}

\subsection{Problem Analysis}
\label{Analysis}
Because we only care about the total cost of all robots, and every coordination requires a robot pair, we can reassign the coordination cost $c'$ from the supporter to the receiver in addition to the reduced edge cost $\tilde{c}_{ij}$, without changing the problem.  
So, the cost with coordination in Eqn.~\eqref{IndividualCost} becomes

\vspace{-10pt}
\begin{align}
\label{NewIndividualCost}
    C_n^t=\begin{cases}
        c_{ij}, ~~\textit{if} ~~ s_{nm}^t=~~0;\\
        \hat{c}_{ij},~~\textit{if} ~~ s_{nm}^t=~~1;\\
        0,~~~~\textit{if} ~~ s_{nm}^t=-1.
    \end{cases}
\end{align} 
where $\hat{c}_{ij}=\tilde{c}_{ij}+c'$. 
Therefore, the original objective function (Eqn.~\eqref{objective}) can be rewritten as
\begin{gather}
    \min_{{\mathcal{M}},{\mathcal{S}}} \sum_{t=0}^{T-1} \sum_{n=1}^N \sum_{\forall i\neq j}M_{ij}^{nt}[(1-s_{nm}^t)c_{ij}+s_{nm}^t \hat{c}_{ij}], 
    \label{new_objective}\\
    s.t. ~~
    \eqref{constraint1:supportnum}, \eqref{constraint2:supportpair}, \eqref{constraint3:starts}, \eqref{constraint4:goals}, 
    \eqref{constraint5:movenum}, \eqref{constraint6:correlation}.\nonumber
\end{gather}

\vspace{-5pt}
\noindent Notice that because when $s_{nm}^t=-1$, $M_{ij}^{nt}=0,\forall i\neq j$, the last \textit{if} condition in Eqn.~\eqref{NewIndividualCost} does not need to be considered in Eqn.~\eqref{new_objective}. To solve the combinatorial optimization problem, a typical approach is dynamic programming~\cite{bellman1966dynamic}, by decoupling the interdependency among the decision variables. The ideal case is to break down the problem into two sub-problems: one optimizing the movement decisions ${\mathcal{M}}$ and the other optimizing the coordination decisions ${\mathcal{S}}$.

Based on such a motivation, if we can find a way to eliminate $n$ from $M_{ij}^{nt}$, Eqn.~\eqref{new_objective} can be rewritten as 
\vspace{-3pt}
\begin{gather}
    \underbrace{\min_{\mathcal{M}} \sum_{t=0}^{T-1} \sum_{\forall i\neq j}M_{ij}^{nt}}_{\textrm{Sub-Problem 2}} \underbrace{\min_{\mathcal{S}} \sum_{n=1}^N [(1-s_{nm}^t)c_{ij}+s_{nm}^t \hat{c}_{ij}]}_{\textrm{Sub-Problem 1}}.
    \label{new_objective2}
\end{gather}
The first half of the function contains only the movement decisions $\mathcal{M}$, while the second half only has the coordination decisions $\mathcal{S}$. Then, decomposition is possible and the NP-hard \textsc{tcgre} problem can be solved with significantly reduced complexity.

%% file: Contents/solution.tex
\section{Solution Algorithms}
\label{sec::solutions}
Based on the mathematical analysis, we propose three classes of algorithms to solve \textsc{tcgre} from different perspectives with different optimality and efficiency characteristics. 
We first propose a class of \textsc{jsg}-based solutions that utilizes the decomposition of the original problem and provides optimal solutions to the two sub-problems (Eqn.~\ref{new_objective2}). Second, to reduce the time complexity, we focus on coordination (i.e., $\mathcal{S}$) and propose a class of coordination-based solutions that decomposes the problem differently. To be specific, we introduce Coordination-Exhaustive Search (\textsc{ces}), which can achieve optimal solutions under a reasonable assumption that each coordination behavior (i.e., support pair composed of support node and risky edge) is only needed for a limited number of times in the optimal solution. Finally, when the same support may need to be repeated many times and the global optimality does not need to be guaranteed, a class of algorithms that focuses only on local sub-team coordination behaviors are introduced, for which we develop Receding-Horizon Optimistic Cooperative A* (\textsc{rhoc-a*}). 

\vspace{-5pt}
\subsection{\textsc{jsg}-Based Solutions}
\label{JSGbased}
\vspace{-3pt}
\textsc{jsg}-based solutions perform the decomposition into the two sub-problems by 1) implicitly solving $\mathcal{S}$ by calculating the minimum edge cost for each edge in the \textsc{jsg}; 2) explicitly solving $\mathcal{M}$ by solving a single-robot shortest path problem with $\mathcal{S}$ implicitly encoded. As mentioned in Sec.~\ref{Analysis}, $n$ is effectively eliminated from $M_{ij}^{nt}$ by building the \textsc{jsg}.

\subsubsection{\textsc{jsg} Construction}
\label{JSG}
In the action model (Sec.~\ref{CostModel}), we use $l_n^t\in \nodeset$ to represent robot $n$'s location at time $t$. In a joint state graph, however, one state is the set of all robots' locations ${\mathbf{L}}^t=\{l_1^t,l_2^t,...,l_N^t\}$. The new node set is $\mathbb{L}=\mathbb{V}^N$ and  each node ${\mathbf{L}_p} \in \mathbb{L}$ correlates to $N$ nodes $(V_{i_1}, V_{i_2}, ..., V_{i_n})$ in $\nodeset$. 
By checking the constraints in Eqns.~\eqref{constraint1:supportnum}, \eqref{constraint2:supportpair}, \eqref{constraint5:movenum}, and  \eqref{constraint6:correlation}, for each pair of joint-states, we can form the new edge set $\mathbb{M}\subset \mathbb{L}^2$. Specially, Eqn.~\eqref{constraint6:correlation} assures no self-loops in the \textsc{jsg}.
An action is the move from current state to next state ${\mathbf{M}}^t=({\mathbf{L}}^t,{\mathbf{L}}^{t+1})$, which can also be written as a 0/1 variable, i.e., ${\mathbf{M}}_{pq}^t=\prod_{n=1}^N M_{ij}^{nt}$, where ${\mathbf{e}}_{pq}$ is any edge in the new edge set, and the movement decision set becomes $\mathcal{M}=\{{\mathbf{M}}^t|\forall t\}$. 
\subsubsection{Sub-Problem 1}
After the construction of \textsc{jsg}, the second half of Eqn.~\eqref{new_objective2} is calculating the minimum edge cost for each $\mathbf{e}_{pq}\in \mathbb{M}$ by optimizing $\mathcal{S}$, which is a 0/1 Integer Linear Programming (ILP) problem and can be solved by classical methods, such as Branch and Bound~\cite{lawler1966branch}.

\subsubsection{Sub-Problem 2}
After solving sub-problem 1, we have the cost $C_{pq}$ for each edge ${\mathbf{e}}_{pq}\in \mathbb{M}$. Sup-problem 2 is to optimize movement decisions $\mathcal{M}$ to minimize the total cost: 
\begin{align}
    \min_{{\mathcal{M}}} ~~\sum_{t=0}^{T-1}\sum_{\forall {\mathbf{e}}_{pq}\in \mathbb{M}} &{\mathbf{M}}_{pq}^t C_{pq}^t, 
    \label{subobjective2}
\end{align}

\vspace{-3pt}
\noindent Now it is a single-robot shortest path problem with non-negative costs and no self-loops, solvable by any shortest path algorithms. Instead of using Dijkstra's algorithm on a fully constructed \textsc{jsg} beforehand~\cite{limbu2023team}, we present results with Uniform Cost Search (\textsc{ucs}) and A* (guided by an admissible heuristics assuming all future risky edges will be supported by a teammate) while constructing \textsc{jsg} on the fly, i.e., interleaving partial solutions of the two sub-problems. 



\subsection{Coordination-Based Solutions}
\label{CoordinationBased}
Some \textsc{jsg} edge costs are simply the sum of individual edge costs of all robots, suggesting possible total cost separation into the costs with and without coordination. Thus, while minimizing costs without coordination can be simply solved for individual robots, the second class of algorithms focuses on coordination. Specifically, we present a Coordination-Exhaustive Search (\textsc{ces}) algorithm based on a slightly different and interleaving decomposition of Eqn.~\eqref{new_objective}: 
\begin{gather}
    \min_{{\mathcal{M}},{\mathcal{S}}} 
    \overbrace{\sum_{t=0}^{T-1}  \sum_{n=1}^N\sum_{\forall e_{ij}\in\edgeset}  M_{ij}^{nt}c_{ij}}^{\textrm{Cost without Coordination}}
    -\label{coordinationbasedobjective2}\\
    \underbrace{\sum_{t=0}^{T-1} \sum_{n=1}^N 
    \sum_{\forall e_{ij}\in\edgeset' or}^{\substack{ l_n^t=l_n^{t+1}\in\\\bigcup_{m=1}^N\mathbb{S}_{M_m^t}}}
    \frac{1}{2}M_{ij}^{nt}[(1-s_{nm}^t)c_{ij}+(1+s_{nm}^t) \Delta{c}_{ij}]}_{\textrm{Cost Reduction due to Coordination}},
    \nonumber
\end{gather}
\vspace{2pt}
where $\Delta{c}_{ij} = {c}_{ij}-\hat{c}_{ij}$. When $s_{nm}^t = 1$ (receiving support) the second part reduces the cost in the first part by $\Delta{c}_{ij}$; When $s_{nm}^t = -1$ (providing support), the cost is reduced to zero. 
Inspired by Conflict-Based Search, we can start with finding the individual shortest path for each robot without coordination. Then, we find the coordination behaviors (with some detours) that can cause the maximum cost reduction.

While the first half can be solved individually for each robot (Fig.~\ref{Proof1} left), \textsc{ces} uses an exhaustive search for the second half. 
If an optimal solution requires a coordination behavior between a robot pair, it is equivalent to make the robot pair detour to the support pair (risky edge and support node) from their original individual shortest paths, while other robots remain on their individual shortest paths (Fig.~\ref{Proof1} right). This robot pair's shortest paths are a combination of two path segments---their shortest paths from their starts to the support pair, and their shortest paths from the support pair to their goals. Given certain coordination behaviors, we can solve the shortest path and minimum total cost of all robots for each path segment (Fig.~\ref{Proof1} right), as shown in Alg.~\ref{CES1}.


\textsc{ces} is a coordination-based method through an exhaustive search. Because one support pair may be assigned to the same/different robot pair(s) for infinite times, to conduct an exhaustive search, we assume each support pair can only occur for a fixed number of times. In our implementation, we assume a support pair can only happen once, but it can be easily expanded to a more general case, by repetitively adding the same support pairs to the coordination set.

The \textsc{ces} algorithm is shown in Alg.~\ref{CES2}: In lines 1-2, using any shortest path algorithm, it generates an individual optimal path and cost for each robot, with original edge costs $\mathbf{c}=\{c_{ij}|\forall e_{ij} \in \edgeset\}$ and reduced edge costs $\hat{\mathbf{c}}=\{\hat{c}_{ij}|\forall e_{ij} \in \edgeset\}$, called pessimistic/optimistic paths $\mathbb{P}_1/\mathbb{P}_2$ and total cost $C_1/C_2$.
If $C_1=C_2$, which means no coordination is needed in the optimal solution, then simply return $\mathbb{P}_1$ and $C_1$ (lines 3-4). Else, it starts the scheduling process. In line 5, it generates a coordination set, $\mathbb{CS}$ that contains all coordination behaviors, and $\mathbb{SCS}$ that contains all subsets of $\mathbb{CS}$, to decide which coordination behaviors/support pairs are needed. In line 6, it generates a set of all robot pairs $\mathbb{RP}$ (order in the pair matters since we need to decide which robot moves to the support node/risky edge) to determine which robot pair should be assigned to each support pair. Now, it looks like a Maximum 3D Matching problem as stated in Definition~\ref{3DMatching}, except that the matching problem has one more assumption that one robot pair can be only assigned once as shown in Fig.~\ref{Proof1} middle. For all possible sets of support pairs (line 8), it generates all possible time orders for this support pair set using permutation (line 9). Notice that robots can wait for one another, so the order of each coordination behavior, not necessarily the exact time step, is sufficient. Then, it iterates through every possible support pair order (line 10). There could be $N(N-1)$ possible robot pairs assigned to each support pair, so an N-Fold Cartesian Product is applied to generate all possible sets of support robot pairs (line 11). Then, it explores every set (line 12), where each robot pair in ${\mathbf{SRP}}$ is assigned to each support pair in ${\mathbf{PSCS}}$ with the same index. Thus, in lines 13-18, it adds the risky edge of each support pair to the individual coordination set of the first robot of the robot pair, and the support node of each support pair to the individual coordination set of the second robot of the robot pair. ${\mathbf{ICS}}$ then contains the individual coordination set of all robots. With ${\mathbf{ICS}}$, we use Alg.~\ref{CES1} to calculate the shortest paths and minimum total cost of this assignment in line 19, which is one solution. Last, lines 20-21 record the best solution with minimum total cost.
To sum up, the loop in line 8 decides a subset of support pairs we need for cost reduction. The loop in line 10 picks an order for all support pairs in the subset. The loop in line 12 selects a robot pair for each support pair in the subset. Iterating through the three loops explores every possible solution under the assumption that each support pair can be applied for a constant number of times, making \textsc{ces} optimal. 
\vspace{-8pt}
\begin{algorithm}    \caption{$CostCalculation$ ($\mathbb{G}, \nodeset_0, \nodeset_g, {\mathbf{ICS}}$)}
    \label{CES1}
    \nl ${\mathbb{P}}=[\emptyset]*N$;\\
    \nl $totalcost=0$;\\
    \nl  \For{$n = 1$ to $ N$}{
    \nl \uIf{${\mathbf{ICS}}[n]=\emptyset$}{
        \nl ${\mathbf{P}},C=ShortestPath(\mathbb{G},\nodeset_0[n],\nodeset_g[n],\mathbf{c})$;\\
        \nl ${\mathbb{P}}[n]={\mathbf{P}}$;\\
        \nl $totalcost=totalcost+C$;\\
    }
    \nl \Else{
        \nl\For{$item\in {\mathbf{ICS}}[n]$}{
            \nl $start = \nodeset_0[n]$;\\
            \nl\uIf{$item$ is a risky edge}{
                \nl ${\mathbf{P}},C=ShortestPath(\mathbb{G},start,item[0],\mathbf{c})$;\\
                \nl ${\mathbb{P}}[n].extend({\mathbf{P}}\bigcup item)$;\\
                \nl $totalcost=totalcost+C+\hat{c}_{item}$;\\
                \nl $start=item[1]$;\\
            }
            \nl\ElseIf{$item$ is a support node}{
                \nl ${\mathbf{P}},C=ShortestPath(\mathbb{G},start,item,\mathbf{c})$;\\
                \nl ${\mathbb{P}}[n].extend({\mathbf{P}})$;\\
                \nl $totalcost=totalcost+C$;\\
                \nl $start=item$;\\ 
            }}
            \nl ${\mathbf{P}},C=ShortestPath(\mathbb{G},start,\nodeset_g[n],\mathbf{c})$;\\
                \nl ${\mathbb{P}}[n].extend({\mathbf{P}})$;\\
                \nl $totalcost=totalcost+C$;\\
        
    }
    }
\nl\Return{$\mathbb{P}, totalcost$}
\end{algorithm}
\vspace{-2pt}

\begin{algorithm}    \caption{\textsc{ces} ($\mathbb{G}, \nodeset_0, \nodeset_g$)}
    \label{CES2}
    \nl $\mathbb{P}_1, C_1 =MultipleShortestPath(\mathbb{G},\nodeset_0,\nodeset_g,\mathbf{c})$;\\
    \nl $\mathbb{P}_2, C_2=MultipleShortestPath(\mathbb{G},\nodeset_0,\nodeset_g,\hat{\mathbf{c}})$;\\
    \nl \If {$C_1\leq C_2$}{
        \nl \Return {$\mathbb{P}_1, C_1$};
    }
    \nl Generate subsets $\mathbb{SCS}=\{{\mathbf{SCS}}|\forall {\mathbf{SCS}}\subset\mathbb{CS}\}$ of the coordination set 
    $\mathbb{CS}$;\\
    \nl Generate a set of all robot pairs $\mathbb{RP}$;\\
    \nl $\mathbb{P}_{min}, C_{min} \leftarrow \mathbb{P}_1, C_1$;\\
    \nl\For{${\mathbf{SCS}}\in\mathbb{SCS}$}{
    \nl ${\mathbb{PSCS}} = AllPermutations({\mathbf{SCS}},len({\mathbf{SCS}}))$;\\
    \nl \For {${\mathbf{PSCS}}\in \mathbb{PSCS}$}{
        \nl ${\mathbb{SRP}}= CartesianProduct(\mathbb{RP},len({\mathbf{PSCS}}))$;\\
        \nl \For {${\mathbf{SRP}}\in{\mathbb{SRP}}$}{
            \nl ${\mathbf{ICS}}=[\emptyset]*N$;\\
            \nl \For{$n=1$ to $len(\bf{PSCS})$}{
            
            \nl $SP={\bf{PSCS}}[n]$;\\
            \nl $RP={\bf{SRP}}[n]$;\\
            \nl ${\mathbf{ICS}}[RP[0]].append(SP[0])$;\\
            \nl ${\mathbf{ICS}}[RP[1]].append(SP[1])$;\\
            }
            \nl $\mathbb{P}, C = CostCalculation(\mathbb{G}, \nodeset_0, \nodeset_g, {\mathbf{ICS}})$;\\
            \nl\If{$C<C_{min}$}{
                \nl $\mathbb{P}_{min}, C_{min}\leftarrow\mathbb{P}, C$;
            }
        }
    }
}    
\nl\Return{$\mathbb{P}_{min}, C_{min}$}.
\end{algorithm}
\vspace{-10pt}
There are a total of $O(2^{|\mathbb{CS}|})$ subsets in $\mathbb{SCS}$ . For each subset, there are $O(|\mathbb{CS}|!)$ permutations. For every permutation, there are $O(N^2)$ possible robot pairs for each support pair, $O(N^{2|\mathbb{CS}|})$ assignments in total. Therefore, the number of possible solutions is $O(2^{|\mathbb{CS}|}\cdot N^{2|\mathbb{CS}|}\cdot |\mathbb{CS}|!)=O((2N^2)^{|\mathbb{CS}|}\cdot |\mathbb{CS}|!)$.
For each solution, we run the shortest path algorithm for $O(|\mathbb{CS}|)$ times; each run costs $O(|\edgeset| log (|\nodeset|))$. Therefore, the time complexity of \textsc{ces} is $O((2N^2)^{|\mathbb{CS}|}\cdot |\mathbb{CS}|!\cdot|\edgeset| log (|\nodeset|))$, which is not exponential to the number of robots $N$ anymore. 
Note that the above algorithm is for directed graphs. For an undirected graph, each edge is equivalently two directed edges in a directed graph, so after line 10, there should be an additional loop that iterates through all possible directions of selected risky edges, which won't change the time complexity class.

\subsection{Receding-Horizon Sub-Team Solutions}

\begin{figure}
    \centering
    \includegraphics[width=0.4\textwidth]{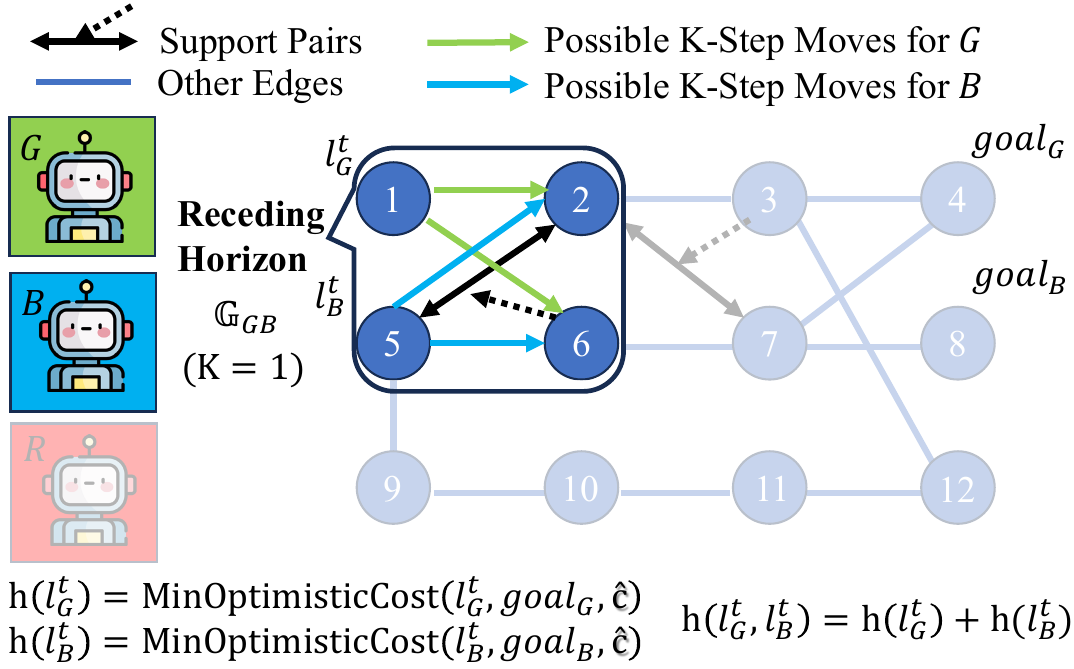}
    \caption{Receding-Horizon Optimistic Cooperative A*.}
    \label{fig::rhoca}
    \vspace{5pt}
\end{figure}

\textsc{tcgre}'s computational complexity arises from two fronts, the large size of both the graph and the team. Therefore, the third class of algorithms reduces the complexity from both fronts by planning with a limited horizon and for a sub-team of all robots at a time, efficiently facilitating local coordination while compromising global optimality. 
Such sub-team local coordination within the receding horizon prioritizes actions that yield the best short-term outcomes and potentially allows dynamic adaptation to changing circumstances, e.g., updated graph structure from robot perception. 
One specific algorithm is Receding-Horizon Optimistic Cooperative A* (\textsc{rhoc-a*}), which provides a flexible and efficient solution by assuring optimal robot pair coordination within the horizon while assuming optimistic cooperation beyond the horizon (Fig.~\ref{fig::rhoca}). 
Alg.~\ref{alg:RHOC-A*} presents \textsc{rhoc-a*}. All robots are initially not at their goals and therefore on duty (lines 1-2). We compute the heuristic for all nodes assuming always-available support (line 3).  \textsc{rhoc-a*} iterates until all robots arrive at their goals (lines 4). If there is still at least one pair of robots on duty (line 5), \textsc{rhoc-a*} sequentially plans for each robot pair (lines 6-7). The computational efficiency is enabled by only looking at a small \textsc{jsg} with only two robots, $n$ and $m$, within $K$ steps (lines 8-12). Notice that optimality is assured on this small \textsc{jsg} while the cost-to-go on the horizon $K$ is estimated by the always-available support heuristic. Lines 13-18 address the situation where only one robot is on duty and has to traverse to its goal alone.

For $N$ robots, generating all possible pairs of robots can be done in $O(N^2)$ time. \textsc{rhoc-a*}'s time complexity for each pair's search of $K$ steps can be approximated as $O(b^K)$, where $b=O(\frac{|\edgeset|}{|\nodeset|})$ is the effective branching factor in the joint action space. 
Thus, in $K$ steps, running \textsc{rhoc-a*} for all pairs results in complexity of $O(N^2(\frac{|\edgeset|}{|\nodeset|})^K)$.
There will be no cycles for any robot, due to  graph search. As a result, there will be at most $O(|V|)$ steps for each robot, so we need to run the $K$-step A* search for $\frac{|V|}{K}$ times for one robot pair. There will be $\frac{N}{2}$ runs for $N$ robots. Therefore, the time complexity of the algorithm is $O(N^3\cdot\frac{|V|}{K}\cdot(\frac{|\edgeset|}{|\nodeset|})^K)$.

\vspace{-5pt}
\begin{algorithm}
\caption{\textsc{rhoc-a*}($\mathbb{G}, \nodeset_0, \nodeset_g$, $K$)}
\label{alg:RHOC-A*}
\nl Initialize $atGoal_n \gets False, \forall n\in \{1,2,...,N\}$\\ 
\nl ${\bf{OnDuty}} = \{n|\forall n, atGoal_n=False\}$\\
\nl Compute optimistic heuristic $h(\cdot)$ for each node in $\mathbb{G}$\\
\nl \While{$len({\bf{OnDuty}})\neq 0$}{
    \nl \If{$len({\bf{OnDuty}})\geq 2$}{
    \nl $\mathbb{RP}=\{\{n,m\}|\forall n,m\in \bf{OnDuty}\}$\\
    \nl\For{each pair $\{n, m\} \in \mathbb{RP}$}{
        \nl Initialize a $K$-step \textsc{jsg} $\mathbb{G}_{nm}$ for the pair $\{n, m\}$ with start $\{l_n^t, l_m^t\}$;\\
        \nl\If{not $atGoal_{n, m}$}{
            \nl A* on $\mathbb{G}_{nm}$ for $K$ steps using $h(\cdot)$;\\
            \nl Update $l_n^t, l_m^t$, $atGoal_{n, m}$, and ${\bf{OnDuty}}$;\\ 
            \nl Update individual and total costs;
        }
    }
    }
    \nl\Else{
        \nl $n={\bf{OnDuty}}.pop()$;\\
        \nl Initialize $K$-step graph $\mathbb{G}_{n}$ with start $l_n^t$;\\
        \nl A* on $\mathbb{G}_{n}$ for $K$ steps using $h(\cdot)$;\\
        \nl Update $l_n^t$, $atGoal_{n}$, and ${\bf{OnDuty}}$;\\ 
        \nl Update individual and total costs;
    }
}
\nl \Return Paths and costs for all robots.

\end{algorithm}

%% file: Contents/results.tex
\vspace{-3pt}
\section{Experiment Results}
\label{sec::results}
\vspace{-3pt}
\definecolor{UCS}{RGB}{128, 0, 129}
\definecolor{A*}{RGB}{9, 5, 255}
\definecolor{CES}{RGB}{244, 1, 4}
\definecolor{RHOC-A*}{RGB}{44, 128, 0}
\definecolor{Naive}{RGB}{241, 16, 255}
\begin{figure*}
    \centering
     \includegraphics[width=0.32\textwidth]{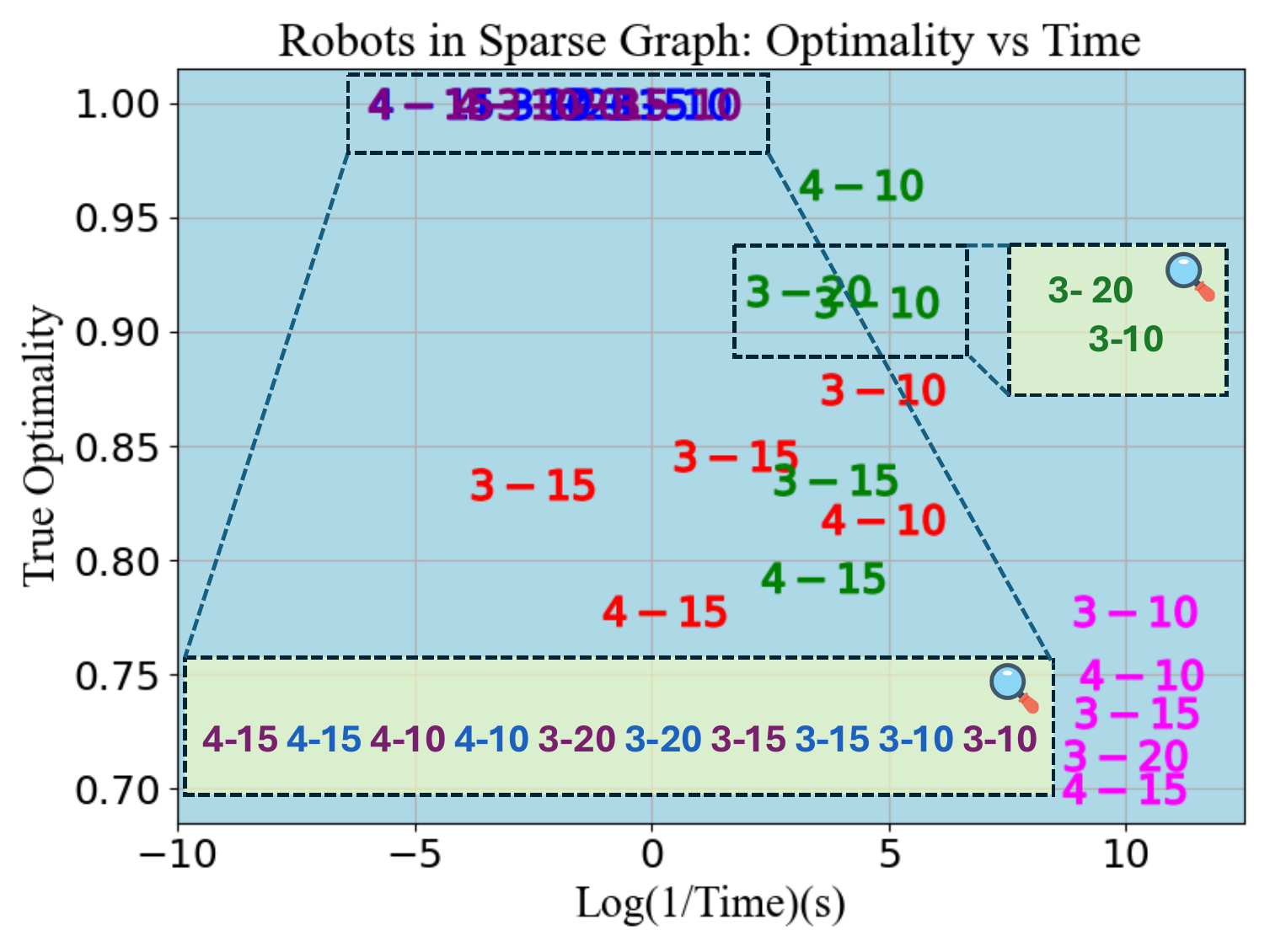}
    \includegraphics[width=0.32\textwidth]{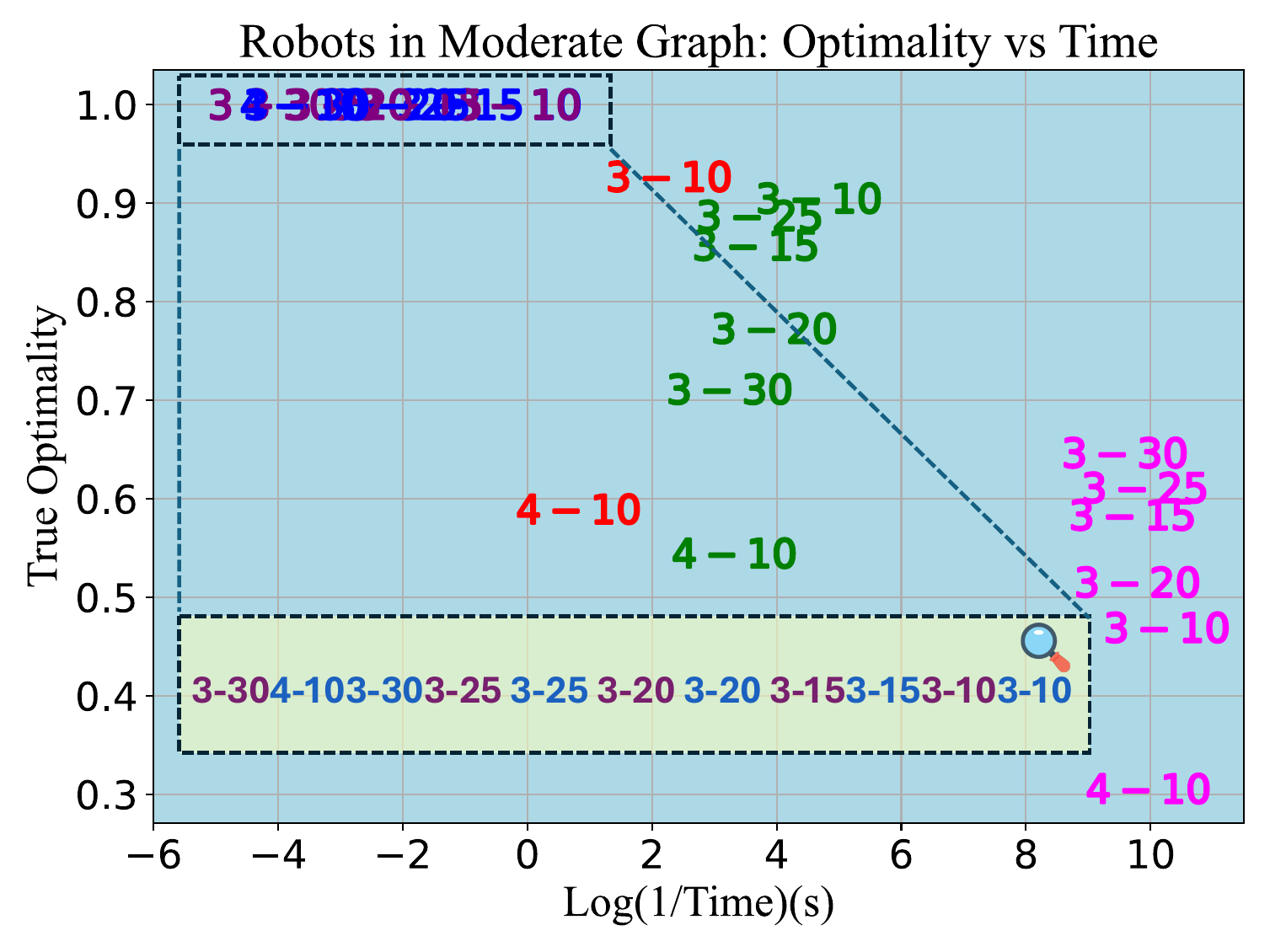}
     \includegraphics[width=0.32\textwidth]{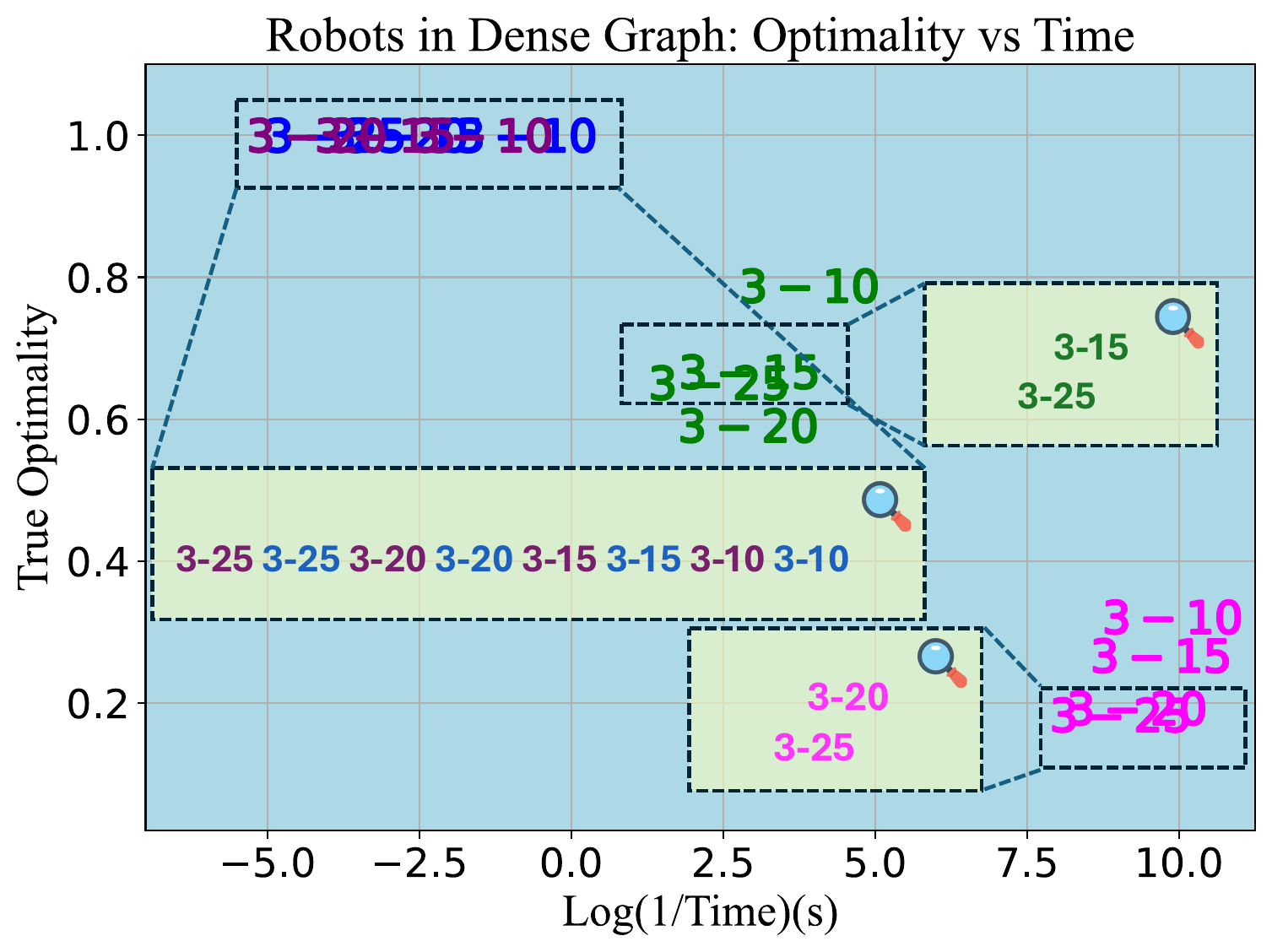}
     \includegraphics[width=0.32\textwidth]{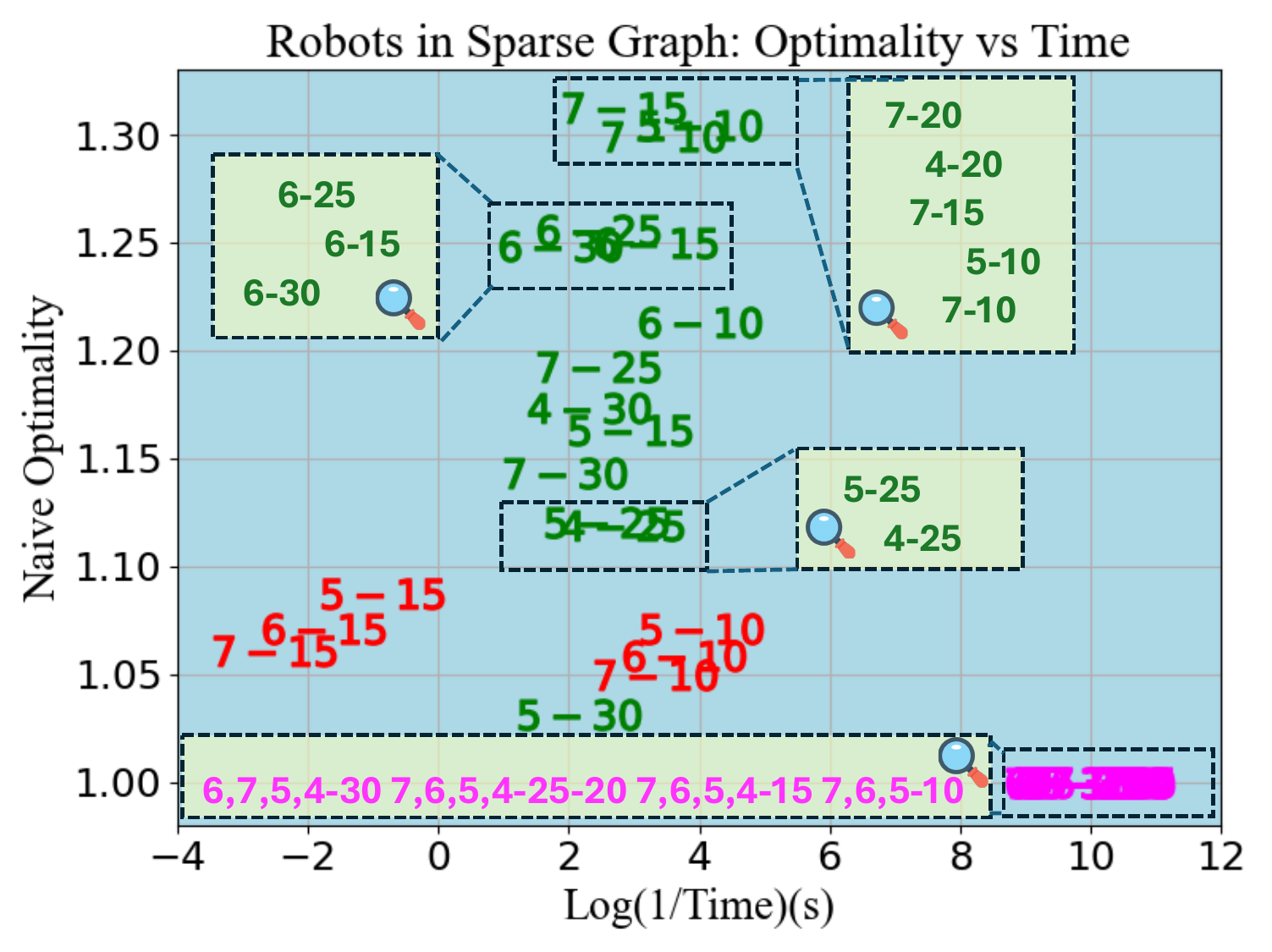}
    \includegraphics[width=0.32\textwidth]{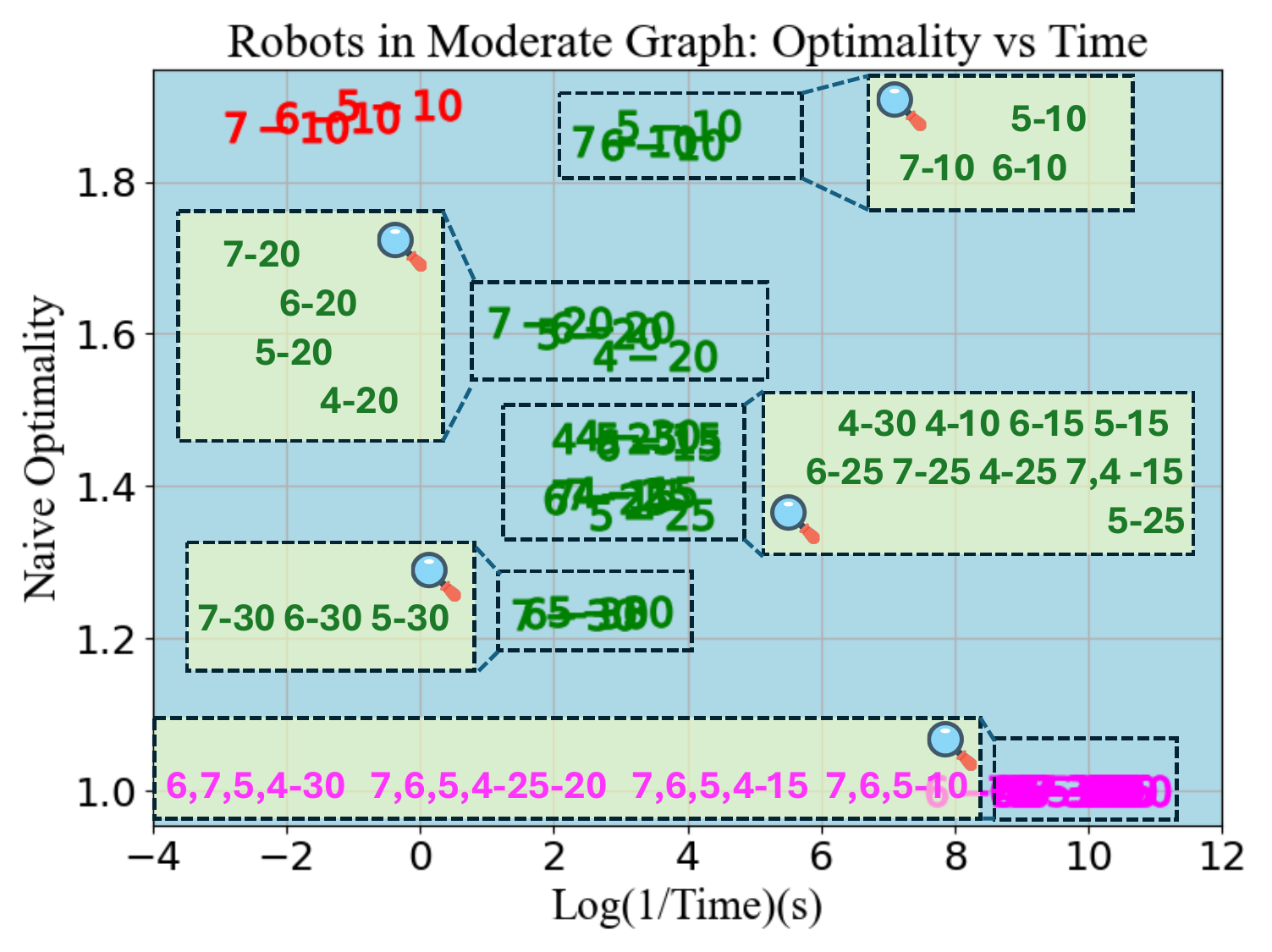}
     \includegraphics[width=0.32\textwidth]{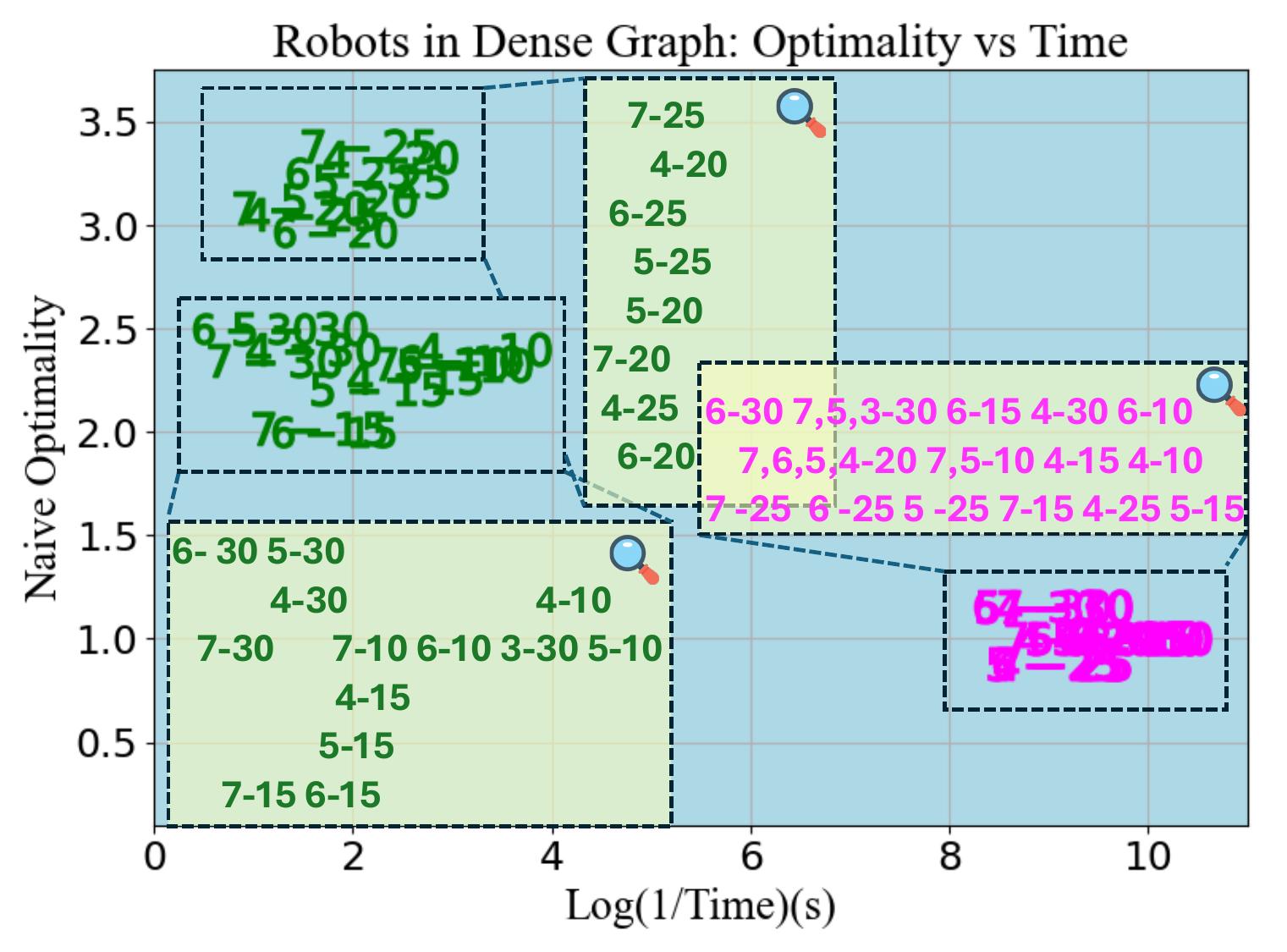}
     \caption{True and Naive Optimality vs. Time with \textcolor{UCS}{\textsc{jsg}-\textsc{ucs}}, \textcolor{A*}{\textsc{jsg}-\textsc{a*}}, \textcolor{CES}{\textsc{ces}}, \textcolor{RHOC-A*}{\textsc{rhoc-a*}}, and \textcolor{Naive}{Naive}. Each data point denotes the result for the experiment with \# of Robots--\# of Nodes. For visibility, cluttered areas are magnified in the dashed boxes. }
     \label{fig:: optimality_on_all_graphs}
     \vspace{-10pt}
\end{figure*}

We conduct experiments on a variety of graphs to evaluate the optimality and efficiency of the three classes of proposed algorithms. To be specific, we implement \textsc{ucs} and \textsc{a*} for \textsc{jsg}-based solutions, \textsc{ces} for coordination-based solutions, and \textsc{rhoc-a*} for receding-horizon sub-team solutions. We first conduct a set of large-scale, method-agnostic experiments on a variety of randomly generated graphs and then present focused experiments to study the pros and cons of specific methods. 

\subsection{Large-Scale Method-Agnostic Experiments}
\vspace{-3pt}
To evaluate each method in an objective manner, we generate a set of graphs with randomly generated support pairs, including sparse, moderate, and dense connectivities and five different numbers of nodes ($|\nodeset| \in \{10, 15, 20, 25, 30\}$), three graphs each type, i.e., a total of 45 distinct graphs. A total of 900 trials are conducted with five different team sizes ($N \in \{3, 4, 5, 6, 7\}$) and four methods. 


We evaluate the optimality and runtime of all methods along with a naive approach, in which each robot executes its individual optimal path without coordination. While the True Optimality value is defined as the optimal cost divided by the actual cost, for scenarios where the optimal cost cannot be found due to excessive computation, we define Naive Optimality to be the naive cost divided by the actual cost. 
If a data point does not exist in Fig.~\ref{fig:: optimality_on_all_graphs}, the corresponding method cannot produce a solution for the robot and node number. 
As shown in Fig.~\ref{fig:: optimality_on_all_graphs}, the \textsc{jsg}-based solutions achieve optimal solutions, but require significant runtime even in small graphs with only a few robots and fail to produce a solution when the the problem becomes larger; \textsc{ces} has better runtime but loses some performance because we assume each support pair can be applied only once; with a fine-tuned $K$, surprisingly, \textsc{rhoc-a*} in many cases achieves better results than \textsc{ces} with less runtime.

\begin{figure}
    \centering
    \includegraphics[width=0.78\columnwidth]{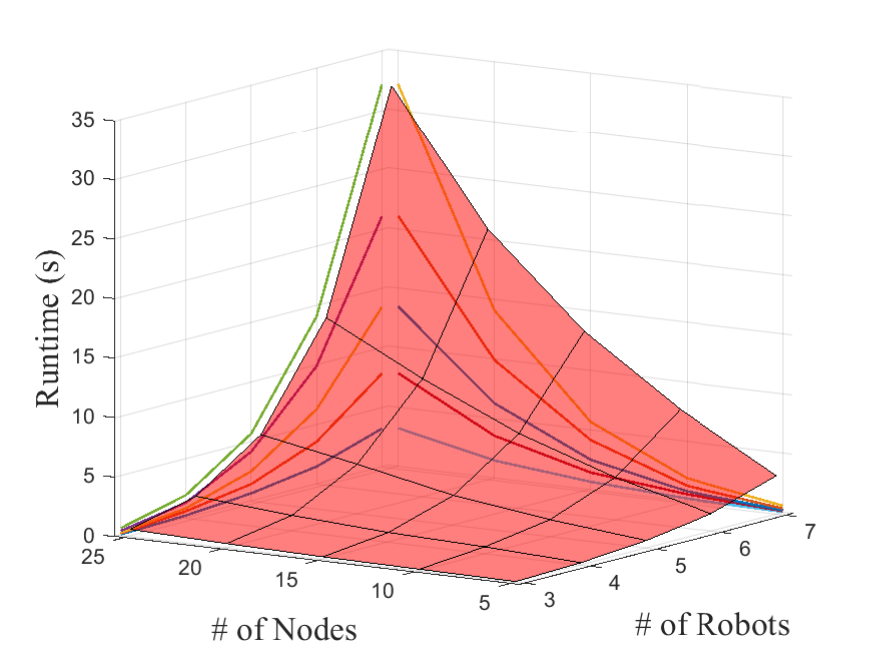}
    \caption{\textsc{ces} Planning Time on Graphs with 2 Support Pairs.}
    \label{fig::ces_3d_plot_2_and_3_support_pairs}
    \vspace{10pt}
\end{figure}

\subsection{Focused Experiments}
\subsubsection{\textsc{ces}'s Insensitivity to Robot and Node Numbers}
Fig.~\ref{fig::ces_3d_plot_2_and_3_support_pairs} showcases that, when there are not many support pairs, \textsc{ces} works well with different numbers of robots and different sizes of graphs (polynomial time to both $N$ and $|\nodeset|$). However, its runtime increases drastically with the number of support pairs, as shown in our method-agnostic experiments, which verifies our time complexity analysis  (Sec.~\ref{CoordinationBased}). 

\subsubsection{\textsc{rhoc-a*}'s Sensitivity to Planning Horizon}
\begin{figure}
    \centering
    \includegraphics[width=0.78\columnwidth]{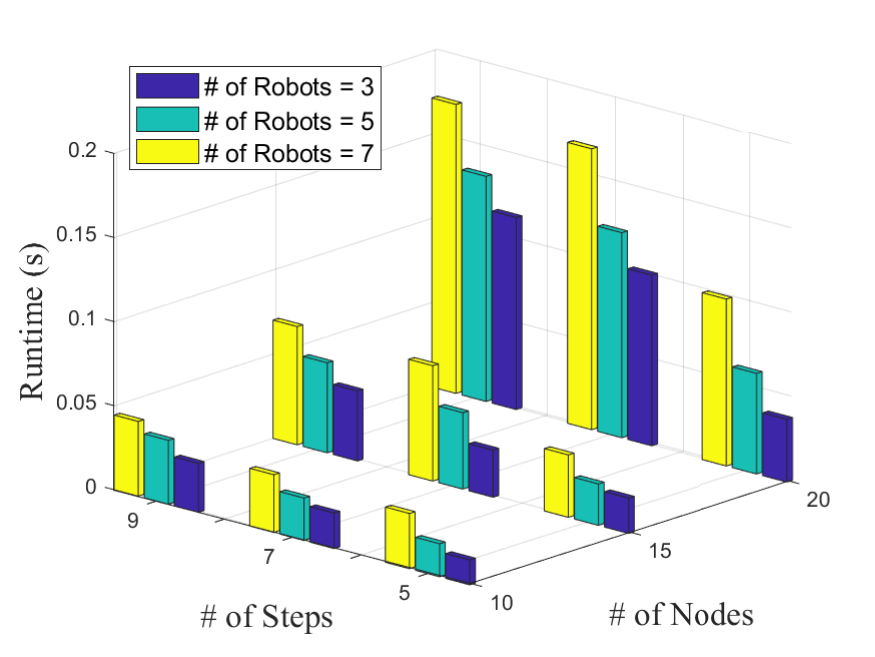}
    \caption{\textsc{rhoc-a*} Planning Time.}
    \label{fig::rhoca_3d_plot}
    \vspace{8pt}
\end{figure}

Fig.~\ref{fig::rhoca_3d_plot} showcases how \textsc{rhoc-a*}'s computation time scales with different planning horizons. A large horizon $K$ comes closer to solving the original \textsc{tcgre} problem with multiple robot pairs, which significantly increases the solution time. While the total cost can be reduced with a longer horizon, it is necessary to strike a balance between horizon and efficiency. 

%% file: Contents/conclusions.tex
\section{Conclusions and Discussions}
\label{sec::conclusions}

We present a systematic problem formulation and mathematical analysis of \textsc{tcgre}, proving its NP-hardness and demonstrating that efficient decomposition is key to solving this problem. We propose three classes of solutions with a set of implementations and present their experimental results.

As given by the analysis in Sec.~\ref{Analysis}, all of the proposed solutions are trying to solve a form of a decomposed problem. For example, \textsc{jsg}-based solutions solve a 0/1 ILP problem and a single-agent shortest path problem, after constructing a \textsc{jsg}; coordination-based solutions, like \textsc{ces}, deal with a 3D matching problem embedded with multiple single-agent shortest path problems. 
By applying some approximation methods to the subproblems---for the former, forming only a few edges instead of all feasible edges and calculating approximate edge costs; for the latter, omitting unpromising matchings---we can significantly reduce runtime without sacrificing too much performance. \textsc{rhoc-a*}, though efficient, does not consider the order of the robot pair selection, with which its performance could improve while still maintaining coordination efficiency.

%% file: root.bbl
\begin{thebibliography}{10}

\bibitem{adler2002cooperative}
J.~L. Adler and V.~J. Blue, ``A cooperative multi-agent transportation management and route guidance system,'' {\em Transportation Research Part C: Emerging Technologies}, vol.~10, no.~5-6, pp.~433--454, 2002.

\bibitem{hu2021distributed}
Y.~Hu, M.~Chen, W.~Saad, H.~V. Poor, and S.~Cui, ``Distributed multi-agent meta learning for trajectory design in wireless drone networks,'' {\em IEEE Journal on Selected Areas in Communications}, vol.~39, no.~10, pp.~3177--3192, 2021.

\bibitem{yu2022surprising}
C.~Yu, A.~Velu, E.~Vinitsky, J.~Gao, Y.~Wang, A.~Bayen, and Y.~Wu, ``The surprising effectiveness of ppo in cooperative, multi-agent games,'' 2022.

\bibitem{surynek2010optimization}
P.~Surynek, ``An optimization variant of multi-robot path planning is intractable,'' in {\em Proceedings of the AAAI conference on artificial intelligence}, vol.~24-1, pp.~1261--1263, 2010.

\bibitem{foerster2017stabilising}
J.~Foerster, N.~Nardelli, G.~Farquhar, T.~Afouras, P.~H. Torr, P.~Kohli, and S.~Whiteson, ``Stabilising experience replay for deep multi-agent reinforcement learning,'' in {\em International conference on machine learning}, pp.~1146--1155, PMLR, 2017.

\bibitem{gupta2017cooperative}
J.~K. Gupta, M.~Egorov, and M.~Kochenderfer, ``Cooperative multi-agent control using deep reinforcement learning,'' in {\em Autonomous Agents and Multiagent Systems: AAMAS 2017 Workshops, Best Papers, S{\~a}o Paulo, Brazil, May 8-12, 2017, Revised Selected Papers 16}, pp.~66--83, Springer, 2017.

\bibitem{liu2019task}
M.~Liu, H.~Ma, J.~Li, and S.~Koenig, ``Task and path planning for multi-agent pickup and delivery,'' in {\em Proceedings of the International Joint Conference on Autonomous Agents and Multiagent Systems (AAMAS)}, 2019.

\bibitem{yan2013survey}
Z.~Yan, N.~Jouandeau, and A.~A. Cherif, ``A survey and analysis of multi-robot coordination,'' {\em International Journal of Advanced Robotic Systems}, vol.~10, no.~12, p.~399, 2013.

\bibitem{luna2011efficient}
R.~Luna and K.~E. Bekris, ``Efficient and complete centralized multi-robot path planning,'' in {\em 2011 IEEE/RSJ International Conference on Intelligent Robots and Systems}, pp.~3268--3275, IEEE, 2011.

\bibitem{capitan2013decentralized}
J.~Capitan, M.~T. Spaan, L.~Merino, and A.~Ollero, ``Decentralized multi-robot cooperation with auctioned pomdps,'' {\em The International Journal of Robotics Research}, vol.~32, no.~6, pp.~650--671, 2013.

\bibitem{limbu2023team}
M.~Limbu, Z.~Hu, S.~Oughourli, X.~Wang, X.~Xiao, and D.~Shishika, ``Team coordination on graphs with state-dependent edge costs,'' in {\em 2023 IEEE/RSJ International Conference on Intelligent Robots and Systems (IROS)}, pp.~679--684, IEEE, 2023.

\bibitem{limbuteam}
M.~Limbu, Z.~Hu, X.~Wang, D.~Shishika, and X.~Xiao, ``Team coordination on graphs with reinforcement learning,'' in {\em 2024 IEEE International Conference on Robotics and Automation (ICRA)}, IEEE, 2024.

\bibitem{sharon2015conflict}
G.~Sharon, R.~Stern, A.~Felner, and N.~R. Sturtevant, ``Conflict-based search for optimal multi-agent pathfinding,'' {\em Artificial Intelligence}, vol.~219, pp.~40--66, 2015.

\bibitem{stern2019multi}
R.~Stern, ``Multi-agent path finding--an overview,'' {\em Artificial Intelligence: 5th RAAI Summer School, Dolgoprudny, Russia, July 4--7, 2019, Tutorial Lectures}, pp.~96--115, 2019.

\bibitem{standley2010finding}
T.~Standley, ``Finding optimal solutions to cooperative pathfinding problems,'' in {\em Proceedings of the AAAI Conference on Artificial Intelligence}, vol.~24-1, pp.~173--178, 2010.

\bibitem{felner2017search}
A.~Felner, R.~Stern, S.~Shimony, E.~Boyarski, M.~Goldenberg, G.~Sharon, N.~Sturtevant, G.~Wagner, and P.~Surynek, ``Search-based optimal solvers for the multi-agent pathfinding problem: Summary and challenges,'' in {\em Proceedings of the International Symposium on Combinatorial Search}, vol.~8-1, pp.~29--37, 2017.

\bibitem{ryan2008exploiting}
M.~R. Ryan, ``Exploiting subgraph structure in multi-robot path planning,'' {\em Journal of Artificial Intelligence Research}, vol.~31, pp.~497--542, 2008.

\bibitem{surynek2012towards}
P.~Surynek, ``Towards optimal cooperative path planning in hard setups through satisfiability solving,'' in {\em Pacific Rim international conference on artificial intelligence}, pp.~564--576, Springer, 2012.

\bibitem{surynek2016makespan}
P.~Surynek, ``Makespan optimal solving of cooperative path-finding via reductions to propositional satisfiability,'' {\em arXiv preprint arXiv:1610.05452}, 2016.

\bibitem{yu2013planning}
J.~Yu and S.~M. LaValle, ``Planning optimal paths for multiple robots on graphs,'' in {\em 2013 IEEE International Conference on Robotics and Automation}, pp.~3612--3617, IEEE, 2013.

\bibitem{erdem2013general}
E.~Erdem, D.~Kisa, U.~Oztok, and P.~Sch{\"u}ller, ``A general formal framework for pathfinding problems with multiple agents,'' in {\em Proceedings of the AAAI Conference on Artificial Intelligence}, vol.~27-1, pp.~290--296, 2013.

\bibitem{surynek2016efficient}
P.~Surynek, A.~Felner, R.~Stern, and E.~Boyarski, ``Efficient sat approach to multi-agent path finding under the sum of costs objective,'' in {\em Proceedings of the twenty-second european conference on artificial intelligence}, pp.~810--818, 2016.

\bibitem{bartak2017modeling}
R.~Bart{\'a}k, N.-F. Zhou, R.~Stern, E.~Boyarski, and P.~Surynek, ``Modeling and solving the multi-agent pathfinding problem in picat,'' in {\em 2017 IEEE 29th International Conference on Tools with Artificial Intelligence (ICTAI)}, pp.~959--966, IEEE, 2017.

\bibitem{Kornhauser1984}
D.~Kornhauser, G.~Miller, and P.~Spirakis, ``Coordinating pebble motion on graphs, the diameter of permutation groups, and applications,'' in {\em 25th Annual Symposium onFoundations of Computer Science, 1984.}, pp.~241--250, 1984.

\bibitem{de2014push}
B.~De~Wilde, A.~W. Ter~Mors, and C.~Witteveen, ``Push and rotate: a complete multi-agent pathfinding algorithm,'' {\em Journal of Artificial Intelligence Research}, vol.~51, pp.~443--492, 2014.

\bibitem{surynek2009novel}
P.~Surynek, ``A novel approach to path planning for multiple robots in bi-connected graphs,'' in {\em 2009 IEEE International Conference on Robotics and Automation}, pp.~3613--3619, IEEE, 2009.

\bibitem{holte1996hierarchical}
R.~C. Holte, M.~B. Perez, R.~M. Zimmer, and A.~J. MacDonald, ``Hierarchical a*: Searching abstraction hierarchies efficiently,'' in {\em AAAI/IAAI, Vol. 1}, pp.~530--535, 1996.

\bibitem{korf1990real}
R.~E. Korf, ``Real-time heuristic search,'' {\em Artificial intelligence}, vol.~42, no.~2-3, pp.~189--211, 1990.

\bibitem{goldreich2011finding}
O.~Goldreich, {\em Finding the shortest move-sequence in the graph-generalized 15-puzzle is NP-hard}.
\newblock Springer, 2011.

\bibitem{kann1991maximum}
V.~Kann, ``Maximum bounded 3-dimensional matching is max snp-complete,'' {\em Information Processing Letters}, vol.~37, no.~1, pp.~27--35, 1991.

\bibitem{bellman1966dynamic}
R.~Bellman, ``Dynamic programming,'' {\em Science}, vol.~153, no.~3731, pp.~34--37, 1966.

\bibitem{lawler1966branch}
E.~L. Lawler and D.~E. Wood, ``Branch-and-bound methods: A survey,'' {\em Operations research}, vol.~14, no.~4, pp.~699--719, 1966.

\end{thebibliography}
